\documentclass[12pt]{iopart}

\usepackage{amsfonts}
\usepackage{graphicx}

\DeclareMathSymbol{\N}{\mathalpha}{AMSb}{'116}
\DeclareMathSymbol{\Z}{\mathalpha}{AMSb}{'132}

\begin{document}

\title{Sudden change of the thermal contact between two quantum systems}
\author{J. Restrepo and S. Camalet}
\address{Laboratoire de Physique Th\'eorique de la Mati\`ere Condens\'ee, 
UMR 7600, Universit\'e Pierre et Marie Curie, Jussieu, Paris-75005, France}
\ead{custserv@iop.org}
\begin{abstract}
In this paper, we address the issue of the stability of the thermal equilibrium of large 
quantum systems with respect to variations of the thermal contact between them. 
We study the Schr\"odinger time evolution of a free bosonic field in two coupled 
one-dimensional cavities after a sudden change of the contact between the cavities. 
Though the coupling we consider is thermodynamically small, modifying it has a 
considerable impact on the two-point correlation functions of the system. We find 
that they do not return to equilibrium but essentially oscillate with a period 
proportional to the length of the cavities. We compare this coupled cavities system 
with the perfect gas which is described by similar expressions but behaves very 
differently.
\end{abstract} 

\pacs{03.65.-w, 05.70.-a, 05.30.Jp}

\maketitle

\section{Introduction}

The physically relevant degrees of freedom of a large isolated system initially 
out of equilibrium, are expected to relax to their thermal 
equilibrium values. Obviously, no relaxation behavior can be observed for 
the complete quantum state of the system but interesting degrees of freedom 
can evolve irreversibly in the limit of a large system. This issue of thermalisation 
in isolated quantum systems has been principally investigated by considering 
composite systems consisting of two distinguishable subsystems of very different 
sizes. For such systems, the smaller subsystem relaxes to thermal equilibrium 
if the larger one is assumed to be in an equilibrium mixed state \cite{QDS}. 
Recently, it has been shown that a priori thermal averaging is not essential. 
The small subsystem can thermalise whereas the whole isolated system is in 
a pure quantum state \cite{EPJB}. 

An isolated system of identical particles cannot be divided into distinguishable 
subsystems and the relevant degrees of freedom are, in this case, the n-particle 
reduced distribution functions such as the particle number density. Relaxation 
behaviors have been obtained for both integrable \cite{IR,SPS,C,RYDO} and 
nonintegrable systems \cite{KLA,MWNM,RDO}. In these studies, the system 
is supposed to be initially in the ground state of a given Hamiltonian but 
its subsequent time evolution is governed by a different Hamiltonian. 
For example, a gas is considered to be initially confined in a subvolume of 
a larger accessible space. The single-particle distributions were always found to 
relax to asymptotic profiles but which are, intriguingly, not all well described by 
an equilibrium ensemble. 

To better apprehend the thermalisation process in isolated many-body systems, 
it is necessary not to be limited to ground states and to be able to treat also finite 
initial temperatures. As we are concerned with isolated systems, such an initial 
condition must be represented by a microcanonical mixed state, according to 
the postulate of equal a priori probabilities, or, equivalently, by a pure state of 
macroscopically well-defined energy \cite{EPJB,PRL,PRE}. In Ref.\cite{PRE}, 
the time-dependent particle density of a perfect quantum gas, initially in such 
a state, has been determined. It was found that this density relaxes from 
a thermal profile to a non-thermal one. In view of these results, one may 
wonder how two quantum systems, initially independent of each other but 
at the same temperature, evolve if they are brought into thermal contact. 
Do their relevant degrees of freedom remain at equilibrium once they are 
coupled ?   
   
In this paper, we address this issue by considering a free bosonic field 
in two one-dimensional cavities coupled to each other by one of their ends. 
Free fields are commonly used to describe the large number of 
environmental degrees of freedom which give rise to dissipation 
in a smaller system \cite{QDS,UZ}. Here, we are interested in two systems 
of comparable size, neither of them acts as a heat bath for the other one. 
In section \ref{M}, we present our model Hamiltonian of two quantum systems 
in thermal contact. This model allows to describe from perfectly coupled to 
completely uncoupled systems. We show that the considered coupling between 
the two cavities is thermodynamically small by evaluating the equilibrium total 
energy and correlation functions. In section \ref{Scc}, we determine how 
the energy contents of the cavities and the field-field correlations evolve in 
response to a sudden change of the coupling between the two cavities. 
Our model is simple enough to obtain exact results for arbitrary values 
of the coupling strength. In section \ref{Cpg}, we discuss why, though 
the expressions are formally similar, relaxation behaviors are obtained for 
a perfect gas but not for the system studied here. Finally, in the last section, 
we summarize our results and mention a few open questions raised by our work.    

\section{Model \label{M}}

\subsection{Derivation of the Hamiltonian \label{dH}}

To describe two coupled one-dimensional bosonic systems characterised by 
length $L$ and velocity $c$, we start from the Hamiltonian 
\begin{equation}
H' = \frac{1}{2} \int_{-L}^{L} dx \left[ \Pi(x)^2 
+ \frac{c^2}{n(x)^2} (\partial_x \phi)^2 \right] 
\label{Hp}
\end{equation}
where the fields $\Pi$ and $\phi$ are canonically conjugate to each other. 
The position-dependent refractive index $n(x)$ is equal to $n>1$ for 
$-d<x<d$ and to $1$ elsewhere. Diverse physical systems can be represented 
by this Hamiltonian. The fields $\Pi$ and $\partial_x \phi$ can be interpreted as 
the electric and magnetic components of the electromagnetic field, or as 
the charge and current distributions of an LC transmission line \cite{QDS,EPL}. 
We write these fields as 
\begin{eqnarray}
\Pi(x) &=& P + i\sqrt{\frac{c}{2}} \sum_{q>0} e^{-q\Lambda/2} \sqrt{q}  
\varphi_q(x)\left( a^{\dag}_q - a^{\phantom{\dag}}_q \right) \label{Piexp} \\
\partial_x \phi &=& \frac{1}{\sqrt{2c}} \sum_{q>0} e^{-q\Lambda/2}  
\frac{\partial_x \varphi_q}{\sqrt{q}} 
\left( a^{\dag}_q + a^{\phantom{\dag}}_q \right) \label{phiexp}
\end{eqnarray}
where the operators $P$ and $a_q$ satisfy the commutation relations 
$[P,a_q]=[a_q,a_{q'}]=0$ and 
$[a^{\phantom{\dag}}_q,a^{\dag}_{q'}]=\delta_{q{q'}}$, and $\Lambda$ is 
a cut-off length. Throughout this paper, we use units in which $\hbar=k_B=1$. 
The eigenmodes $\varphi_q$ of $H'$ are solutions of the differential equation 
$-q^2 \varphi_q=\partial_x [\partial_x \varphi_q / n(x)^2]$ with the boundary 
conditions $\partial_x \varphi_q (-L)=\partial_x \varphi_q (L)=0$. The odd 
$ \varphi_q$ are proportional to $\mathrm{sgn} (x) \cos [q(|x|-L)]$ for $|x|>d$, 
where the wavenumber $q$ obeys 
\begin{equation}
\tan[q(L-d)]^{-1}=n\tan(2nqd) .  \label{eqq0}
\end{equation}
The even eigenmodes are proportional to $\cos [q(|x|-L)]$ for $|x|>d$, with $q$ 
given by $\tan[q(L-d)]=-n\tan(2nqd)$. 

To obtain the Hamiltonian of two cavities coupled to each other by one of 
their ends, we consider the limits $n \gg 1$ and $d \ll L$ with the length 
\begin{equation}
D = n^2 d
\end{equation}
kept fixed. In this regime, equation \eref{eqq0} simplifies to $\tan(qL)=2/Dq$ 
which can be rewritten as
\begin{equation}
\frac{Dq+2i}{Dq-2i} e^{-2iqL} = 1 . \label{eqq}
\end{equation}
This second form will be useful in the following. For the even eigenmodes, 
the wavenumbers become $q=p\pi/L$ where $p$ is a positive integer. It can 
then be shown, using \eref{Piexp} and \eref{phiexp}, that the Hamiltonian 
\eref{Hp} turns into
\begin{equation}
H = H_\mathrm{left} + H_\mathrm{coupling} + H_\mathrm{right}  \label{H}
\end{equation}
where 
\begin{equation}
H_\mathrm{left} = \frac{1}{2} \int_{-L}^0 dx 
\left[ \Pi(x)^2 + c^2 (\partial_x \phi)^2 \right] ,
\label{Hl}
\end{equation}
$H_\mathrm{right}$ is given by this expression with $-L$ and $0$ replaced 
by $0$ and $L$, respectively, and
\begin{equation}
H_\mathrm{coupling} =  \frac{c^2}{2D} \left[ \phi(0^+)-\phi(0^-) \right]^2 . 
\label{Hc}
\end{equation}
In the following, we study the Hamiltonian \eref{H} which allows to describe 
from perfectly coupled to completely uncoupled cavities. Its eigenmodes 
$\varphi_q$ obey $\partial_x \varphi_q \left( 0^- \right)
=\partial_x \varphi_q \left( 0^+ \right)
= \left[  \varphi_q \left( 0^+ \right) - \varphi_q \left( 0^- \right) \right]/D$ which leads 
to even $\varphi_q (x) = L^{-1/2} \cos(qx)$ independent of the coupling 
characteristic length $D$. The odd eigenmodes read
\begin{equation}
\varphi_q (x) = A_q \mathrm{sgn} (x) \cos [q(|x|-L)] \label{oe}
\end{equation}
where $A_q=[L+(D/2)(1+(Dq/2)^2)^{-1}]^{-1/2}$ and $q$ is a positive solution 
of \eref{eqq}. In the large $D$ limit, the two cavities decouple from each other as 
can be seen from the fact that the boundary conditions at the ends $0^-$ and $0^+$ 
become similar to that at the ends $-L$ and $L$. In this limit, the solutions $q$ of 
the equation \eref{eqq} are the multiples of $\pi/L$ and the corresponding eigenmodes 
$\varphi_q $ simplify to $ L^{-1/2} \mathrm{sgn} (x) \cos(qx)$ for $q>0$ and to 
$ (2L)^{-1/2} \mathrm{sgn} (x)$ for the solution $q \rightarrow 0$. In the opposite limit, 
$D=0$, the odd eigenmodes are given by $ \varphi_q (x) = L^{-1/2} \sin(qx)$ with 
the wavenumbers $q=(p+1/2)\pi/L$ where $p \in \N$. The transmission at $x=0$ is 
perfect in this case. 

\subsection{Thermal contact \label{Tc}}

\begin{figure}
\centering \includegraphics[width=0.65\textwidth]{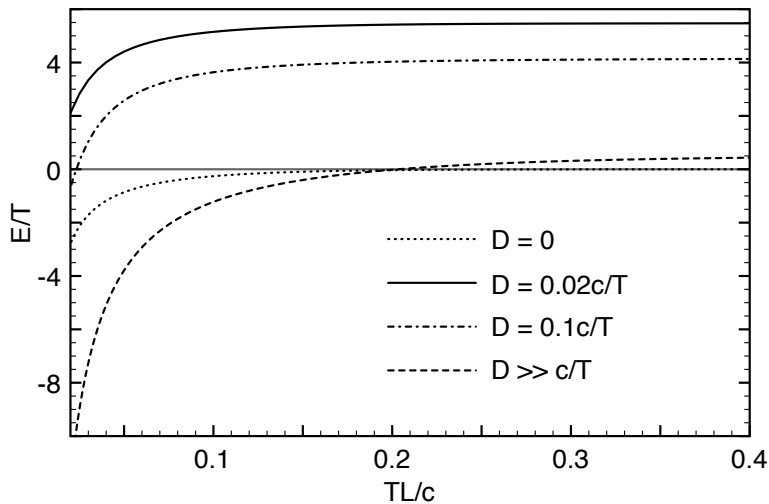}
\caption{Energy $E$ as a function of length $L$ for $\Lambda=0.01c/T$ and 
different values of $D$. $E$ reaches a maximum at $D \simeq 0.03c/T$. 
The corresponding curve is indistinguishable from that obtained for $D=0.02c/T$.  
The results for $D > 100c/T$ and $D < 10^{-4}c/T$ are well-described by 
the large and vanishing $D$ approximations \eref{Einf} and \eref{E0}, respectively. 
\label{fig:E}}
\end{figure}

The coupling \eref{Hc} is thermodynamically small, i.e., its contributions to 
the thermodynamic functions of the complete system described by \eref{H} are 
negligible in the thermodynamic limit of large $L$ and fixed energy density. 
As an example, we consider the average energy 
$\langle H \rangle_T=\mathrm{Tr}[\exp(-H/T)H]/\mathrm{Tr}[\exp(-H/T)]$ 
at temperature $T$. 
Diagonalizing the Hamiltonian \eref{H} with the help of the expansions 
\eref{Piexp} and \eref{phiexp} gives 
\begin{equation}
\langle H \rangle_T = \frac{T}{2} 
+ \sum_{q>0} \frac{cq e^{-q\Lambda}}{2\tanh(cq/2T)}  . 
\label{HT}
\end{equation}
As there is only one solution to equation \eref{eqq} in a given interval 
$[p\pi/L,(p+1)\pi/L]$ where $p \in \N$, the above sum over $q$ can be 
approximated by an integral, which leads to
\begin{equation}
\langle H \rangle_T =  \frac{L c}{\pi\Lambda^2} + \frac{\pi}{3c} L T^2
+E \left( T D/c, TL/c \right) . \label{HTint}
\end{equation} 
The first two extensive terms stem from the integral approximation, see \ref{AA}. 
The energy $E$ is the difference between this approximation and the sum \eref{HT} 
and is thus expected to be negligible in the limit $L \gg c/T$. For large and 
vanishing $D$, the wavenumbers $q$ are equally spaced and it is then possible 
to evaluate $E$, see \ref{AA}. We find
\begin{eqnarray}
E \left( \infty , TL/c \right) &=& \frac{T}{2} 
-  \frac{2\pi}{c} T^2 L \sum_{p>0} \sinh(2p\pi TL/c)^{-2} 
\label{Einf} \\
E \left( 0 , TL/c \right) &=& 
-  \frac{2\pi}{c} T^2 L \sum_{p>0} \sinh(4p\pi TL/c)^{-2} . \label{E0}
\end{eqnarray} 
Note that the energy $\langle H \rangle_T$ for large $D$ is not exactly twice 
the energy of an isolated cavity of length $L$, the coupling between the two cavities 
contributes an energy $T/2$. In the limit $L \gg c/T$, the energy \eref{Einf} reaches 
$T/2$ and \eref{E0} vanishes. For an arbitrary characteristic coupling length $D$, 
$E/T$ converges also to a finite value in this limit, as shown by Fig.~\ref{fig:E}, 
and hence the energy of the complete system 
$\langle H \rangle_T \simeq L c/\pi\Lambda^2+ \pi L T^2/3c$ is essentially 
independent of $D$. We remark that, since the solutions of \eref{eqq} are 
$q \simeq p\pi/L + 2/\pi  p D$ where $p$ is a positive integer, for large $D$, $E$ first 
increases with decreasing $D$. The expressions \eref{Einf} and \eref{E0} are valid 
for any $LT/c$. We see that, for $L \ll c/T$, the contribution of the coupling \eref{Hc} 
cannot be neglected as $E \left( \infty , TL/c \right)$ and  $E \left( 0 , TL/c \right)$ 
diverge differently in this limit. We obtain this diverging behavior of the energy $E$ 
for other values of $D$ as well, see Fig.~\ref{fig:E}.

\subsection{Equilibrium correlation functions}

\begin{figure}
\centering \includegraphics[width=0.65\textwidth]{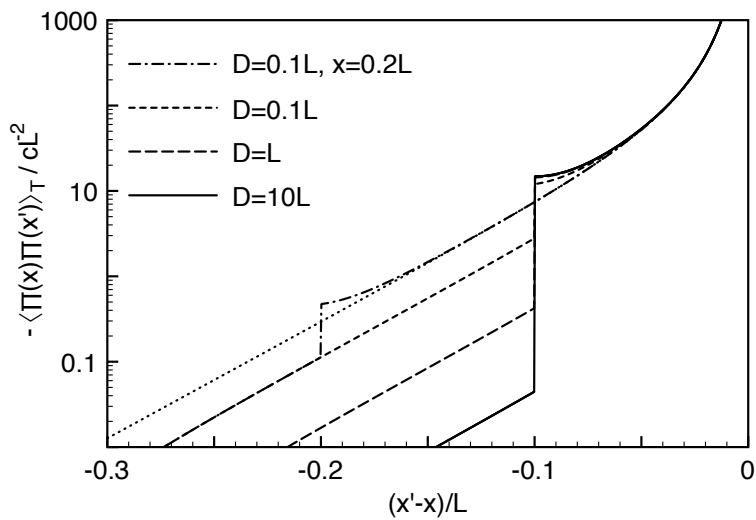}
\caption{Equilibrium correlation function $\langle \Pi(x) \Pi(x') \rangle_T$ as 
a function of $x'-x$ for $T=5c/L$, $x=0.1L$ and $0.2L$, and $D=0.1L$, $L$ and 
$10L$. The dotted line corresponds to the thermodynamic limit expression 
\eref{corrTTL}. For the above parameters, the agreement with this approximation 
is excellent for $x'>x$.  
\label{fig:correq}}
\end{figure}

It is also instructive to consider the thermal two-point correlation functions of 
the system, for example 
\begin{eqnarray}
&\langle \Pi(x) \Pi(x') \rangle_T = \frac{c}{8L} \sum_{q=p\pi/L} 
\frac{q e^{-\Lambda |q|}}{\tanh(c q/ 2T) }  \left( e^{iq(x-x')} + e^{iq(x+x')} \right) 
\label{corrT2} \\
&\qquad + \frac{c}{8} \mathrm{sgn} (xx') \sum_q 
\frac{A_q^2 q e^{-\Lambda |q|}}{\tanh(c q/ 2T) } 
 \left( e^{iq(|x|-|x'|)} + \frac{Dq-2i}{Dq+2i} e^{iq(|x|+|x'|)} \right) \nonumber
\end{eqnarray}
where $p$ runs over all integers and the second sum over the negative and 
positive solutions of \eref{eqq}. For vanishing and large $D$, this expression 
can be simplified by using the relation \eref{rel} and the function \eref{G}. 
We find, for $D=0$ and $D \rightarrow \infty$, respectively, 
\begin{eqnarray}
\langle \Pi(x) \Pi(x') \rangle^0_T &=& -\frac{\pi T^2}{2c} 
\sum_{p,\epsilon} \sinh^{-2}  
\left[\frac{\pi T}{c}\Big( x+\epsilon x'+(4p+1+\epsilon)L \Big) \right]
\label{PiPi0} \\
\langle \Pi(x) \Pi(x') \rangle^\infty_T &=& -\frac{\pi T^2}{2c} \Theta(xx')
\sum_{p,\epsilon} \sinh^{-2} 
\left[\frac{\pi T}{c}\Big( x+\epsilon x'+2pL \Big) \right] \label{PiPiinf} 
\end{eqnarray}
where $p$ runs over $\Z$ and $\epsilon$ over $\{ -1,1 \}$. For $L \gg c/T$, 
the only terms which contribute significantly to the sum \eref{PiPi0} are 
$(\epsilon,p)=(-1,0)$, $(1,-1)$ and $(1,0)$. The last two are important  
only for $(x,x')$ in the vicinity of $(-L,-L)$ and $(L,L)$. The vanishing of 
\eref{PiPiinf} for $xx'<0$ shows that there is no correlation between the two 
cavities in the large $D$ limit. The comparison of \eref{PiPi0} and \eref{PiPiinf} 
reveals that, in the large $D$ case, the thermal correlation function \eref{corrT2} 
is identical to that of two isolated cavities of length $L$. Consequently, in these 
two limiting cases, 
\begin{equation}
\langle \Pi(x) \Pi(x') \rangle_T \simeq -\frac{\pi T^2}{2c} \sinh^{-2}[\pi T(x-x')/c] 
\label{corrTTL}
\end{equation}
for $x$ and $x'$ not too close to the cavities ends. More precisely, 
$\langle \Pi(x) \Pi(x') \rangle_T$ deviates notably from this approximate 
expression if $x$ or $x'$ is at a distance smaller than $c/T$ from $-L$, $0$ 
or $L$. For other values of $D$, the field-field correlations \eref{corrT2} 
are also well described, in the large $L$ limit, by \eref{corrTTL}. This can 
be seen as follows. For $x \simeq x'$ not too close to the ends $-L$, $0$ 
and $L$, the terms $\exp[iq(x+x')]$ and $\exp[iq(|x|+|x'|)]$ in \eref{corrT2} 
vary a lot from one value of $q$ to the next and hence give vanishing 
contributions. Moreover, the normalisation factor $A_q$ is close to 
$L^{-1/2}$ except possibly for the first few roots of \eref{eqq}, and 
this equation has only one solution in a given interval 
$[p\pi/L,(p+1)\pi/L]$ where $p \in \N$. As a result, the two remaining sums 
in \eref{corrT2} are accurately approximated by the same integral which 
leads to \eref{corrTTL}. The thermal correlation function \eref{corrT2} 
differs significantly from this approximation only for $x$ and $x'$ close 
to the cavities ends as shown in Fig.~\ref{fig:correq}.

For the field $c\partial_x \phi$, we obtain the expression
\begin{eqnarray}
&c^2 \langle \partial_x \phi (x) \partial_{x'} \phi (x') \rangle_T 
= \frac{c}{8L} \sum_{q=p\pi/L} 
\frac{q e^{-\Lambda |q|}}{\tanh(c q/ 2T) }  \left( e^{iq(x-x')} - e^{iq(x+x')} \right) 
\label{cdxphicdxphi} \\
&\qquad + \frac{c}{8} \sum_q \frac{A_q^2 q e^{-\Lambda |q|}}{\tanh(c q/ 2T) } 
 \left( e^{iq(|x|-|x'|)} - \frac{Dq-2i}{Dq+2i} e^{iq(|x|+|x'|)} \right) \nonumber
\end{eqnarray}
where $p$ runs over $\Z$ and the second sum over the solutions of \eref{eqq}. 
For $D=0$ and $D \rightarrow \infty$, this correlation function is given by \eref{PiPi0} 
and \eref{PiPiinf}, respectively, with an extra factor $-\epsilon$ in the summand. 
For other values of $D$, the above arguments can be used to simplify 
\eref{cdxphicdxphi}. Thus, the approximate expression \eref{corrTTL} applies to 
the field $c\partial_x \phi$ as well. For the correlations between the fields $\Pi$ 
and $\partial_x \phi$, the completeness of the basis $\{ \varphi_q \}$ 
implies, for any $D$, $\langle \partial_x \phi (x) \Pi(x')  \rangle = (i/2) \delta'(x-x')$. 
In summary, in the thermodynamic limit, the thermal correlation functions are 
essentially independent of the characteristic coupling length $D$. 

\section{Sudden change of the coupling strength \label{Scc}}

We consider here that the system is initially at equilibrium with temperature $T$ 
and characteristic coupling length $D_0$ and then evolves under the Hamiltonian 
\eref{H} with $D\ne D_0$. We study the time-dependent two-point correlation 
functions and the time evolution of the energy contained in one cavity. Since, for 
these expectation values, a microcanonical mixed state or a pure state of 
macroscopically well-defined energy are, in the thermodynamic limit, equivalent to 
a canonical ensemble \cite{EPJB}, we evaluate canonical averages in the following. 

\subsection{Time-evolved field operators}

\begin{figure}
\centering \includegraphics[width=0.65\textwidth]{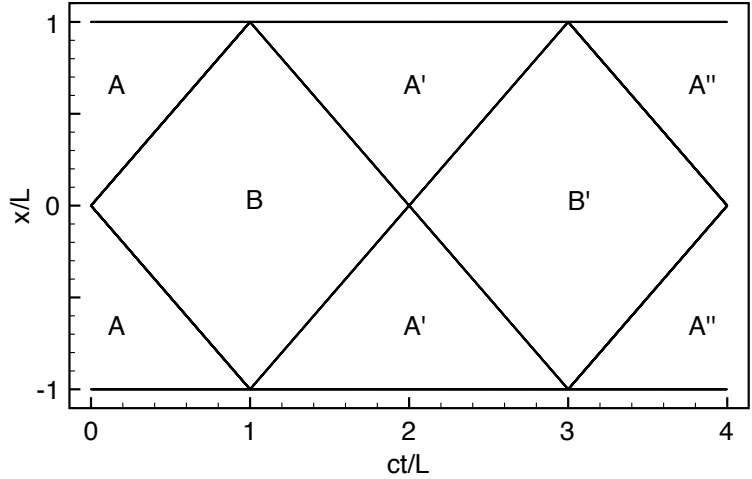}
\caption{The time-dependence of the fields $\Pi$ and $\partial_x \phi$ is different 
in space-time regions $A$, $B$, \ldots, see the text. \label{fig:figexp} }
\end{figure}

To obtain the field-field correlations at any time $t$, we write 
the time-evolved field operators in terms of the creation operators 
$a^\dag_{k}$ corresponding to $D_0$ as
\begin{eqnarray}
\partial_x \phi (x,t) &=& e^{iHt} \partial_x \phi e^{-iHt} 
= \frac{1}{\sqrt{2c}} \sum_{k > 0}
e^{-k\Lambda/2} \left[ \Theta_{k}(x,t) a_k^\dag + \Theta^*_{k}(x,t) 
a_k^{\phantom{\dag}} \right]  \label{dxphi} \\
\Pi (x,t) &=&  P + i\sqrt{\frac{c}{2}} \sum_{k > 0}
e^{-k\Lambda/2} \left[ \Omega_{k}(x,t) a_k^\dag - \Omega^*_{k}(x,t) 
a_k^{\phantom{\dag}} \right]  . \label{Pi} 
\end{eqnarray}
The coefficients $\Theta_{k}$ and $\Omega_k$ are given by 
\begin{eqnarray}
\Theta_{k}(x,t) &=& k^{-1/2} \sum_{q>0} (k|q)
\partial_x \varphi_q \left[  \cos(cqt) + i \frac{k}{q} \sin(cqt) \right]  \label{Thetak} \\
\Omega_{k}(x,t) &=& k^{1/2} \sum_{q>0} (k|q)
\varphi_q(x) \left[  \cos(cqt) + i \frac{q}{k} \sin(cqt) \right]  \label{Omegak}
\end{eqnarray}
where $(k|q)=\int_{-L}^L dx \varphi^{(0)}_{k} (x)\varphi_{q} (x)$, $\varphi_{q}$ 
and $\varphi^{(0)}_{k}$ are the eigenmodes corresponding to $D$ and $D_0$, 
respectively. If $\varphi^{(0)}_{k}$ is an even function of $x$ then only the even 
$\varphi_{q}$ contribute to the sums \eref{Thetak} and \eref{Omegak}. Moreover, 
in this case, since the even eigenmodes do not depend on the characteristic 
coupling length, these expressions simplify to 
$\Theta_k= -(k/L)^{1/2} \sin(kx) \exp(ickt)$ 
and $\Omega_k= (k/L)^{1/2} \cos(kx) \exp(ickt)$. For odd $\varphi^{(0)}_{k}$, it 
is instructive to first consider the two limiting cases of vanishing and large $D$.

\subsubsection{Perfectly coupled cavities}

For $D=0$, the odd eigenmodes are $\varphi_q (x)=L^{-1/2} \sin(qx)$ with 
the wavenumbers $q=(p+1/2)\pi/L$ where $p \in \N$. By evaluating $(k|q)$ and 
using the relation \eref{rel}, we obtain 
\begin{eqnarray}
\Omega_{k} &=& 2iLA_k \cos(kL) k^{-1/2} \sum_p (-1)^p 
\Big[ \delta (\tau-{\bar x}-p) - \delta (\tau+{\bar x}-p) \Big] \nonumber \\
&~&-\frac{A_k}{2} k^{1/2} e^{-ikL} \Big[ (-1)^{\lfloor \tau-{\bar x} \rfloor} 
e^{2ikL \epsilon(\tau-{\bar x})}
 - (-1)^{\lfloor \tau+{\bar x} \rfloor} e^{2ikL \epsilon(\tau+{\bar x})} \Big] 
 \label{OmegakD0}
\end{eqnarray}
where $p$ runs over all integers, $\tau=ct/2L$, ${\bar x}=x/2L$, 
$\lfloor \tau \rfloor$ denotes the largest integer smaller than $\tau$, and 
$\epsilon(\tau)=\tau-\lfloor \tau \rfloor$. Inspection of this expression shows that 
the field $\Pi$ is the superposition of two waves travelling with velocity $c$ and 
reflected with no phase shift at $x=\pm L$. For a given $x$, $|\Omega_{k}|$ takes 
two different values as time goes on, see Fig \ref{fig:figexp}. We find 
\begin{eqnarray}
\frac{e^{-ikct} \Omega_{k}}{A_k \sqrt{k}}  &= e^{i(\pi-2kL)p} 
\mathrm{sgn} (x) \cos[k(|x|-L)] & 
\hbox{  in regions A, A', \ldots} \nonumber \\
&=  e^{i(\pi-2kL)(p+1/2)} \sin(kx) 
& \hbox{  in regions B, B', \ldots}  \label{eOmegak}
\end{eqnarray}
where $p=0$ in $A$/$B$, $1$ in $A'$/$B'$, \ldots . The space-time regions 
$A$, $B$, $A'$, $B'$, \ldots do not depend on the wavenumber $k$. As we will see 
below, this independence plays an essential role in the behavior of physical 
properties such as the field-field correlations or the cavities energies. 
For the field $\partial_x \phi$, the coefficient \eref{Thetak} is given by a similar 
expression.

\subsubsection{Uncoupled cavities}

For large $D$, we obtain
\begin{equation}
\Omega_{k} = \mathrm{sgn}(x)\frac{A_k}{2} k^{1/2} e^{-ikL} 
\Big[ e^{2ikL \epsilon(\tau-{\bar x})} + e^{2ikL \epsilon(\tau+{\bar x})} \Big] .
\label{OmegakDinf}
\end{equation}
Here, the field $\Pi$ is the superposition of two waves reflected at $x=0$ 
and $L$ ($-L$) in the right (left) cavity. 
The expression \eref{eOmegak} becomes 
\begin{eqnarray}
\frac{e^{-ikct} \Omega_{k}}{A_k \sqrt{k}}  &= e^{-2ikLp} 
\mathrm{sgn} (x) \cos[k(|x|-L)] & 
\hbox{  in regions A, A', \ldots} \nonumber \\
&= e^{-2ikL(p+1/2)}
\mathrm{sgn} (x) \cos(kx) & \hbox{  in regions B, B', \ldots}  
\label{eOmegakinf}
\end{eqnarray}
where $p=0$ in $A$/$B$, $1$ in $A'$/$B'$, \ldots . Contrary to the above case, 
the field $\Pi$ is always discontinuous at $x=0$. The field $\partial_x \phi$ 
satisfies $\partial_x \phi(0,t)=0$, i.e., the reflection is perfect at $x=0^+$ and $0^-$.

\subsubsection{General case}

For a finite value of $D$, the fields $\Pi$ and $\partial_x \phi$ are also 
superpositions of propagating waves but which are, contrary to the above studied 
special cases, both transmitted and reflected at $x=0$. We write the coefficients 
$\Omega_k$ and $\Theta_k$ as
\begin{eqnarray}
\Omega_k (x,t) &=& \mathrm{sgn} (x) \left( \frac{D}{D_0} -1 \right) 
A_k \cos (kL) k^{-1/2} \left[ \Gamma_k^+(x,t) + \Gamma_k^-(x,t) \right] 
\label{Omegakgen} \\
\Theta_k (x,t) &=& i \left( \frac{D}{D_0} -1 \right) 
A_k \cos (kL) k^{-1/2} \left[ \Gamma_k^+(x,t) - \Gamma_k^-(x,t) \right] 
\label{Thetakgen} \\
\Gamma_k^\pm(x,t) &=&  \sum_q \frac{q  A_q^2}{k-q}  \frac{e^{iq(ct\pm |x|)}}
{Dq\pm 2i} \label{Gammadef}
\end{eqnarray}
where the sum runs over the positive and negative solutions of \eref{eqq}. 
To evaluate the functions $\Gamma_k^\pm$, we first note that, since $q$ is 
a solution of \eref{eqq}, the summand in \eref{Gammadef} can be multiplied 
by a factor $[(Dq+2i) \exp(-2iqL)/(Dq-2i)]^m$ where $m$ is any integer. 
We then rewrite $\Gamma_k^\pm$ as  
\begin{eqnarray}
\Gamma_k^\pm (x,t) &=& \frac{1}{\pi D} \oint_{\cal C} dz 
\frac{X(z)^{m+1}}{1-X(z)}\frac{ze^{iz(ct\pm |x|)/D}}{(Dk-z)(z\pm 2i)} \nonumber \\ 
&~&+2i\frac{X(Dk)^{m+1} }{1-X(Dk)} \frac{k e^{ik(ct \pm |x|)}}{Dk \pm 2i}  
\label{Gamma}
\end{eqnarray} 
where $X(z)=(z+2i)\exp(-2izL/D)/(z-2i)$ and $\cal C$ is a contour enclosing 
the real axis. Expressing the above integral in terms of the residues of the integrand 
poles on the real axis leads to \eref{Gammadef}. For times $t$ between $t^\pm_m$ 
and $t^\pm_{m+1}$ where $t^\pm_m=\mp |x|/c+2mL/c$, the upper (lower) part of 
the contour $\cal C$ can be closed in the upper (lower) half plane. Moreover, for 
positive $m$, the integrand in \eref{Gamma} has poles only on the real axis and 
at $z=2i$. Thus, for $t >0$, the first term of the expression \eref{Gamma} is 
determined by the residue of this last pole if $m$ is chosen as the largest integer 
smaller than $(ct\pm |x|)/2L$. It can be shown that the functional dependence of 
this term on $\tau=c(t-t^\pm_m)/D$ is $\exp(-2\tau)P^\pm_m(\tau)$ where 
$P^+_m$ ($P^-_m$) is a polynomial of order $m-1$ ($m$), see \ref{AB}. 
The functions $\Omega_k$ and $\Theta_k$ change discontinuously at 
$t = t^\pm_m$. For $D \ll L$, they then reach the values given by the second term 
of \eref{Gamma} in a time of order $D/c$. For $D=0$, one retrieves the Dirac delta 
functions of \eref{OmegakD0} and the second term of \eref{Gamma} gives 
the expression \eref{eOmegak}. In the opposite limit, $D \gg L$, the first term 
of \eref{Gamma} is negligible with respect to the second one which leads to 
the expression \eref{eOmegakinf}. Contrary to the limiting cases of vanishing 
and large $D$ studied above, since the successive reflections and transmissions 
at $x=0$ deform the propagating waves, $|\Omega_k|$ and $|\Theta_k|$ are not 
strictly periodic here.

\subsection{Correlation functions}

\begin{figure}
\centering \includegraphics[width=0.65\textwidth]{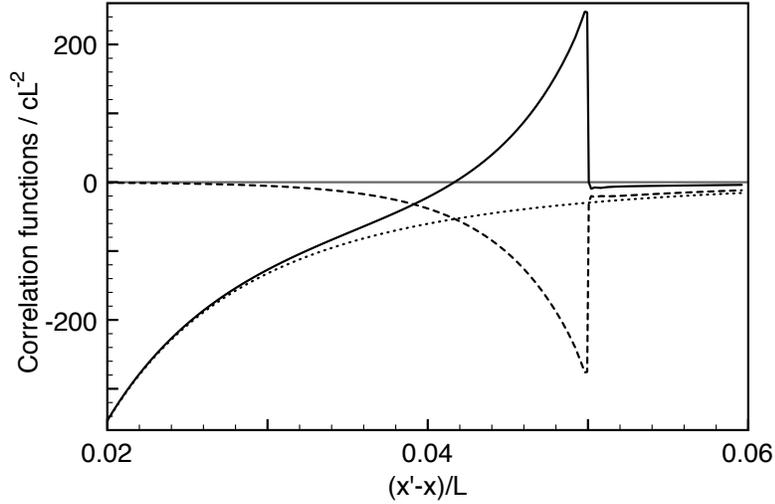}
\caption{\label{fig:corr} Correlation functions $\langle \Pi (x,t) \Pi (x',t) \rangle$
(full line) and $\langle c \partial_x \phi(x,t) \Pi (x',t) \rangle$ (dashed line) as 
functions of $x'-x$ for $x=0.5L$, $T=10c/L$, $D_0=100 L$, $D=0.01L$, 
and $t=0.55 L/c$. The discontinuities correspond to $x'=ct$, see Fig.~\ref{fig:figexp}. 
The spatial correlations of the field $c\partial_x \phi$ are indistinguishable from that 
of the field $\Pi$ (full line). The dotted line is the thermal equilibrium correlation 
function \eref{corrTTL}.}
\end{figure}

\begin{figure}
\centering \includegraphics[width=0.65\textwidth]{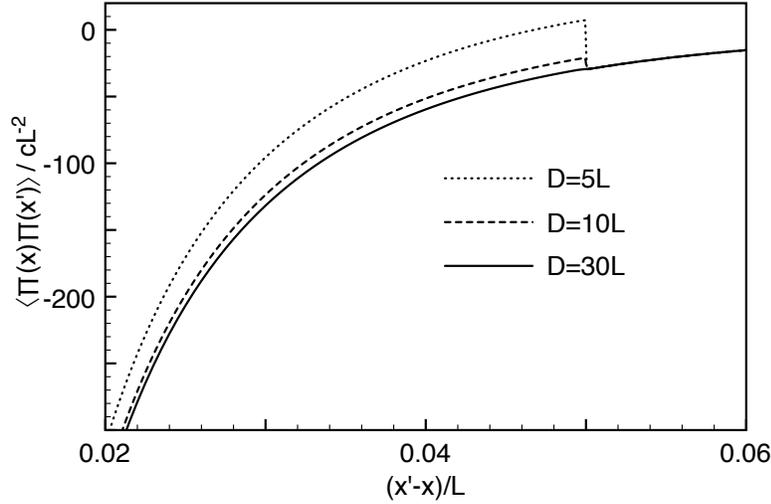}
\caption{\label{fig:corr2} 
Two-point correlation function of the field $\Pi$ for $D=5 L$, $10 L$ and $30 L$. 
The other parameters are as in Fig.\ref{fig:corr}. The curves for $D \ge 30 L$ are 
indistinguishable from one another.}
\end{figure}

With the notations introduced in \eref{dxphi} and \eref{Pi}, the two-point correlation 
functions of the cavities read
\begin{eqnarray}
&\langle \Pi(x,t) \Pi(x',t) \rangle =  \frac{T}{2L} +  \frac{c}{2} \mathrm{Re}
\sum_{k > 0} \frac{\Omega_{k}(x,t) 
\Omega^*_{k}(x',t)}{\tanh(c k/ 2T)} e^{-k\Lambda} \label{corrxxp} \\
&\langle \partial_x \phi (x,t) \partial_{x'} \phi (x',t) \rangle =  \frac{1}{2c} 
\mathrm{Re}\sum_{k > 0} \frac{\Theta_{k}(x,t) \Theta^*_{k}(x',t)}{\tanh(c k/ 2T)} 
e^{-k\Lambda} \label{corrxxp2} \\
&\langle \partial_x \phi (x,t) \Pi(x',t) \rangle =  \frac{1}{2} \mathrm{Im}
\sum_{k > 0} \frac{\Theta_{k}(x,t) \Omega^*_{k}(x',t)}{\tanh(c k/ 2T)} 
e^{-k\Lambda} + \frac{i}{2} \delta' (x-x') . 
\label{corrxxp3}
\end{eqnarray}
Since the functions $\Theta_k$ and $\Omega_k$ present discontinuities at 
positions independent of $k$, as discussed above, the field-field correlations 
at a given time change abruptly at $x=x_m$ and $x'=x_m$ 
where $x_m=2mL\pm ct$, $m \in \Z$, see Fig.~\ref{fig:corr} and \ref{fig:corr2}. 
The results shown in figures \ref{fig:corr} and \ref{fig:corr2} are 
obtained by numerically solving \eref{eqq} and evaluating the exact expressions 
\eref{Thetak}-\eref{Omegak} and \eref{corrxxp}-\eref{corrxxp3}. For small $D$, 
the first term of \eref{Gamma} contributes significantly to the correlations 
\eref{corrxxp}-\eref{corrxxp3} only if $x$ or $x'$ is close to $x_m$, as shown in 
Fig~\ref{fig:corr}. This term becomes negligible in the large $D$ limit, see 
Fig.~\ref{fig:corr2}. Thus, the field-field correlations are essentially described 
by the second term of \eref{Gamma} for small and large $D$. Keeping only 
this term and using the properties $X(Dk)^*=X(Dk)^{-1}=X(-Dk)$, we obtain
\begin{eqnarray}
\langle \Pi(x,t) \Pi(x',t) \rangle &\simeq  \frac{c}{8 L} \sum_{k=p\pi/L} 
\frac{k e^{-\Lambda |k|}}{\tanh(c k/ 2T)}  \left( e^{ik(x-x')}+ e^{ik(x+x')} \right) 
\label{corrfin} \\
& \hspace{-1cm} +\frac{c}{8} \mathrm{sgn} (xx') \sum_{k} 
\frac{A_k^2 k e^{-\Lambda |k|}}{\tanh(c k/ 2T)}  
\left( \alpha_k e^{ik\big| |x|-|x'| \big| }+ \beta_k e^{ik(|x|+|x'|-2L)} \right) 
\nonumber
\end{eqnarray}
where the second sum runs over the positive and negative solutions of 
\eref{eqq} with the characteristic coupling length $D_0$. The coefficients 
$\alpha_k$ and $\beta_k$ are given by
\begin{eqnarray}
(\alpha_k,\beta_k) &= (1,1) & \hbox{  for } (x,x') 
\in A\times A, A'\times A', \ldots \nonumber \\
&= ((1+X)/2,(1+X^*)/2) & \hbox{  for } (x,x') 
\in A\times B, B \times A, A'\times B, \ldots  \nonumber \\
&= (1,X^*) & \hbox{  for } (x,x') \in B\times B, B'\times B', \ldots 
\label{alphabeta}
\end{eqnarray}
where $X=X(Dk)$ and the regions $A$, $B$, $A'$, \ldots are presented 
in Fig.~\ref{fig:figexp}. Therefore, the approximate correlation function 
\eref{corrfin}, for given $x$ and $x'$, is periodic with period $2L/c$. 
When $x$ and $x'$ are both in a region of type $A$, the expression 
\eref{corrfin} is identical to the initial condition \eref{corrT2}. As discussed 
after equation \eref{corrTTL}, the second and fourth terms of \eref{corrfin} 
are negligible except for $x$ and $x'$ near a cavity end, and hence 
\eref{corrfin} is practically equal to \eref{corrT2} for $x$ and $x'$ both in 
a region of type $B$, see Fig.~\ref{fig:corr}. When $x$ and $x'$ are in 
regions of different types, \eref{corrfin} can differ considerably from 
\eref{corrT2}. For example, for cavities perfectly coupled ($D=0$) but 
initially uncoupled ($D_0 \rightarrow \infty$), $\alpha_k=\beta_k=0$ in 
this case and thus the spatial correlations of the field $\Pi$ are reduced 
by a factor of two with respect to the initial condition \eref{corrTTL}. 
It is interesting to note that even for uncoupled cavities 
($D \rightarrow \infty$), the sudden change of the characteristic coupling 
length causes a time-evolution of the correlation functions as 
$X=\exp(-2iLk)$ is equal to $1$ only for $D_0 \rightarrow \infty$. 

For the other correlation functions, we find
\begin{eqnarray}
\langle \partial_x \phi (x,t) \partial_{x'} \phi (x',t) \rangle &\simeq  
\frac{1}{8 c L} \sum_{k=p\pi/L} \frac{k e^{-\Lambda |k|}}{\tanh(c k/ 2T)}  
\left(  e^{ik(x-x')} - e^{ik(x+x')} \right)  \label{corrfin2} \\
& \hspace{-1cm} +\frac{1}{8c} \sum_{k} 
\frac{A_k^2 k e^{-\Lambda |k|}}{\tanh(c k/ 2T)}  
\left( \alpha_k e^{ik\big| |x|-|x'| \big|} - \beta_k e^{ik(|x|+|x'|-2L)}  
\right) \nonumber
\end{eqnarray}
where $\alpha_k$ and $\beta_k$ are given by \eref{alphabeta}, and, 
for $x \ne x'$,
\begin{eqnarray}
\langle \partial_x \phi (x,t) \Pi(x',t) \rangle &\simeq& 
\frac{\mathrm{sgn} (x')}{16} \sum_{k} 
\frac{A_k^2 k e^{-\Lambda |k|}}{\tanh(c k/ 2T)} \label{corrfin3} \\  
&~& \qquad \qquad \times \left( \gamma_k e^{ik\big| |x|-|x'| \big| }
\pm \delta_k e^{ik(|x|+|x'|-2L)} \right) \nonumber
\end{eqnarray}
where the sum runs over the solutions of \eref{eqq}, the upper sign is 
for $|x|>|x'|$, and
\begin{eqnarray}
(\gamma_k,\delta_k) &= (0,0) & \hbox{  for } (x,x') 
\in A\times A, B \times B, A'\times A', \ldots 
\nonumber \\
&= (1-X,X^*-1) & \hbox{  for } (x,x') \in A\times B, B \times A, A'\times B', \ldots  
\nonumber \\
&= (X-1,1-X^*) & \hbox{  for } (x,x') \in A'\times B, B \times A', A''\times B', \ldots
\end{eqnarray}
Contrary to the correlations \eref{corrfin} and \eref{corrfin2} of a field with itself, 
the correlations between the fields $\Pi$ and $\partial_x \phi$ given by \eref{corrfin3} 
vanish when $x$ and $x'$ are in the same region, see Fig.~\ref{fig:corr}. For $x$ 
and $x'$ not too close to a cavity end, 
$\langle \partial_x \phi (x,t) \partial_{x'} \phi (x',t) \rangle \simeq 
\langle \Pi(x,t) \Pi(x',t) \rangle/c^2$ and $\langle \partial_x \phi (x,t) \Pi(x',t) \rangle 
\simeq \pm (\langle \Pi(x,t) \Pi(x',t) \rangle-\langle \Pi(x,0) \Pi(x',0) \rangle)/c$ 
where the sign depends on the relative positions of $0$, $x$ and $x'$. We have 
seen in the previous section that there is no correlation between the two fields 
and that the correlations of a field with itself are essentially the same everywhere 
at equilibrium. Switching the characteristic coupling length to another value 
induces correlations between the two fields and affects the whole system even far 
away from the interface between the two cavities.
 
\subsection{Energy of the cavities \label{Ec}}

\begin{figure}
\centering \includegraphics[width=0.75\textwidth]{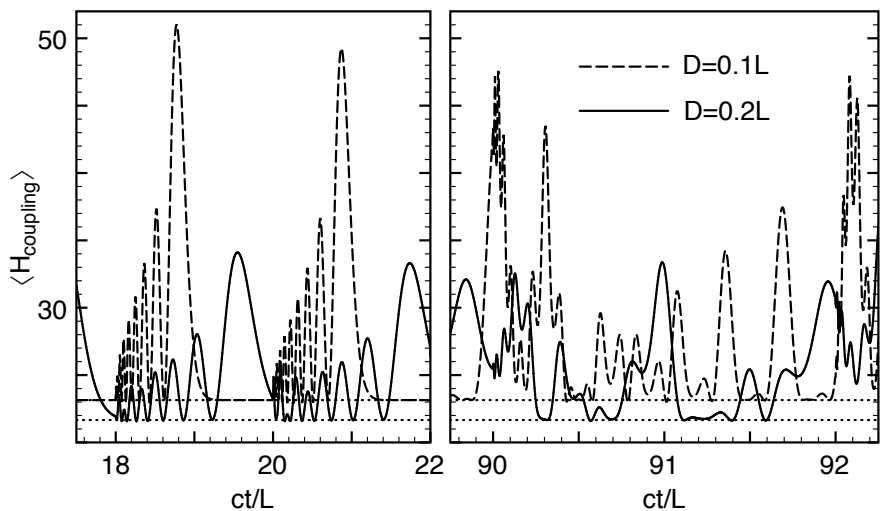}
\caption{Coupling energy as a function of time for $D_0=L$, $T=10c/L$ and 
$D=0.1L$ and $0.2L$. In the left part of the figure, $t<t_0$ for $D=0.1L$ and 
$t \simeq t_0$ for $D=0.2L$. The dotted lines correspond to the small $D$ 
approximation \eref{Hct}. 
\label{fig:Hct}}
\end{figure}
 
We study here the time evolution of the energy content of the cavities. 
For that purpose, we first note that the energy density 
\begin{equation}
\frac{1}{2}\langle \Pi(x,t)^2 \rangle 
+ \frac{c^2}{2} \langle \partial_x \phi (x,t)^2 \rangle 
= \frac{T}{4L} + \frac{c}{4} \sum_{k>0} 
\frac{ |\Omega_k (x,t)|^2 + |\Theta_k (x,t)|^2}{\tanh(ck/2T) e^{k\Lambda}} . 
\label{ed}
\end{equation}
is even with respect to $x$. This symmetry property leads to 
\begin{equation}
\langle H_\mathrm{left} (t) \rangle = \langle H_\mathrm{right} (t) \rangle 
=  \langle H \rangle /2 - \langle H_\mathrm{coupling} (t) \rangle/2 .
\end{equation}
The time dependence of the cavities energies is thus simply related to 
that of the coupling energy. As the eigenmodes of the total Hamiltonian 
$H$ satisfy $\partial_x \varphi_q \left( 0^- \right)=\partial_x \varphi_q \left( 0^+ \right)= 
\left[  \varphi_q \left( 0^+ \right) - \varphi_q \left( 0^- \right) \right]/D$, the time-evolved 
coupling Hamiltonian \eref{Hc} is proportional to $\partial_x \phi(0,t)$ and hence 
can be expressed in terms of the functions \eref{Thetak}. We obtain
\begin{equation}
\langle H_\mathrm{coupling} (t) \rangle = \frac{c}{4} D \sum_{k>0} 
\frac{ |\Theta_k (0,t)|^2  e^{-k\Lambda}}{\tanh(ck/2T)} . \label{Hct0}
\end{equation}
The even eigenmodes do not contribute to this sum since their derivatives vanish 
at $x=0$. For the odd eigenmodes, we have seen above that, for $D \ll L$, 
the first term of the expression \eref{Gamma} contributes only for $t$ close to 
$2mL/c$ where $m$ is an integer. The second term of \eref{Gamma} gives, 
using $|X(Dk)|=1$,
\begin{equation}
\langle H_\mathrm{coupling} (t) \rangle \simeq c D \sum_{k>0} 
\frac{A_k^2 k  e^{-k\Lambda}}{\tanh(ck/2T)} \frac{1}{4+(Dk)^2} \label{Hct}
\end{equation}
where the sum runs over the wavenumbers $k$ corresponding to odd 
$\varphi_k^{(0)}$. For a small but finite $D$, the coupling energy deviates from 
this value only for short periods of time when $t$ is much smaller than 
a characteristic time $t_0$ which increases with decreasing $D$. In this short 
time regime, the energy density \eref{ed} is essentially constant except in two 
regions wich propagate with velocity $c$ and are reflected and transmitted at 
$x=0$, $L$ and $-L$. The multiple peak structure shown in left part of 
Fig.~\ref{fig:Hct} results from the preceding reflections and transmissions 
at $x=0$. The temporal evolution of $\langle H_\mathrm{coupling} (t) \rangle$ 
becomes more complicated for $t>t_0$, see Fig.~\ref{fig:Hct}. For $D \gg L$, 
the contributions of the first and second terms of \eref{Gamma} to the functions 
$\Theta_k(0,t)$ are of the same order and hence the approximate expression 
\eref{Hct} does not apply. For $D \rightarrow \infty$, $\partial_x \phi(0,t)$ 
vanishes, as mentioned after \eref{eOmegakinf}. However, because of 
the factor $D$ in \eref{Hct0}, some care must be taken in the evaluation 
of $\langle H_\mathrm{coupling} (t) \rangle$ in this limit. For large $D$, 
the wavenumbers $q$ are practically the multiples of $\pi/L$ except the lowest 
one $q \simeq (2/DL)^{1/2}$. The terms of the sum \eref{Thetak} are thus 
essentially equal to their infinite $D$ values except the first one and 
\begin{equation}
\langle H_\mathrm{coupling} (t) \rangle \simeq  
\sin^2\left( \frac{ct}{\sqrt{DL/2}}  \right) \frac{c}{2L} 
\sum_{k>0} \frac{  A_k^2 \sin^2(kL) e^{-k\Lambda}}{k \tanh(ck/2T)} 
\end{equation}
varies sinusoidally with a finite amplitude.

\section{Comparison to perfect gas \label{Cpg}}

\begin{figure}
\centering \includegraphics[width=0.65\textwidth]{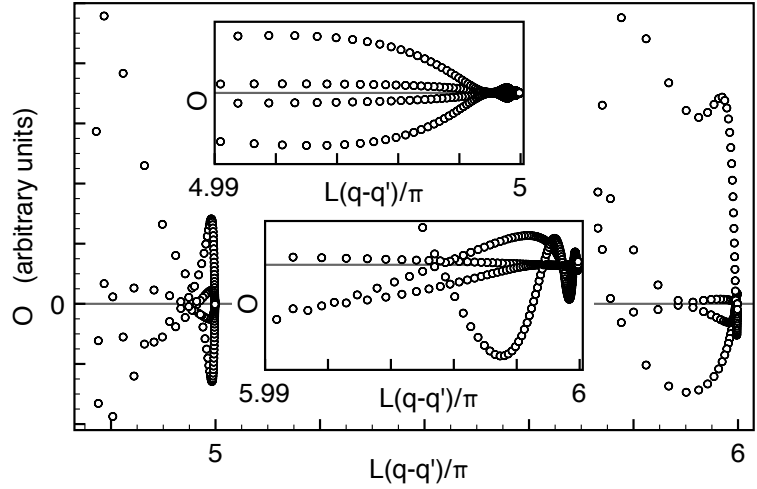}
\caption{Coefficient $O_{qq'}$ as a function of frequency 
$\omega_{qq'}=c(q-q')$ for the correlation function 
$\langle \Pi(x,t) \Pi(x',t) \rangle$ with 
$x=0.5 L$, $x'=0.51 L$, $D_0=0$, $D=0.1 L$ and $T= 10 c/L$. 
\label{fig:spectre}}
\end{figure}

The time-dependent average energies $\langle H_\mathrm{left} (t) \rangle$, 
$\langle H_\mathrm{right} (t) \rangle$ and 
$\langle H_\mathrm{coupling} (t) \rangle$, 
and the correlation functions \eref{corrxxp}-\eref{corrxxp3} can be written as 
\begin{equation}
\langle \hat O (t) \rangle = \sum_{q,q'} O_{qq'} e^{it(\epsilon_{q}-\epsilon_{q'})} 
= \int d\omega e^{it \omega} O(\omega) \label{Ot}
\end{equation}
where $\hat O$ stands for $H_\mathrm{left}$, \ldots, the sums run over 
the positive and negative $q$ corresponding to the characteristic coupling length 
$D$, $O_{qq'}$ are appropriate coefficients, $\epsilon_q=cq$ and 
$O(\omega)=\sum_{q,q'>0} O_{qq'} \delta(\omega-\epsilon_q+\epsilon_{q'})$. 
We remark that the even eigenmodes $\varphi_q$ contribute only to the steady 
component of $\langle \hat O (t) \rangle$, see the discussion after equation 
\eref{Omegak}. The expectation values of the single-particle observables of 
an isolated perfect gas, its particle number density at a given point for instance, 
are also given by expressions of the form \eref{Ot} with the single-particle energies 
$\epsilon_q$ \cite{PRL}. Consider as an example a one-dimensional perfect gas 
confined in a box of length $L$. In this case, the single-particle energies are 
$\epsilon_q = p^2 \pi^2/2mL^2$ where $m$ is the particle mass and $p$ an integer. 
The frequencies $\epsilon_q-\epsilon_{q'}$ are thus regularly spaced and 
$\langle \hat O (t) \rangle$ is periodic with period $4mL^2/\pi$. However, in 
the Joule expansion studied in Ref.~\cite{PRL}, for times $t \ll mL^2$, the function 
$O$ in the integral expression \eref{Ot} of the gas density profile, can be 
approximated as a continuous function plus a term $O_\infty \delta(\omega)$ and 
$\langle \hat O (t) \rangle$ essentially relaxes from its initial value to $O_\infty$.   
 
The situation is radically different for the free field system considered in this paper. 
There obviously also exists a $\delta(\omega)$ contribution to $O(\omega)$ but 
$\langle \hat O (t) \rangle$ does not relax. First of all, we observe that, in the limiting 
cases of vanishing and large $D$, the function $O$ is a sum of equally spaced Dirac 
delta functions and $\langle \hat O (t) \rangle$ is strictly periodic with period $2L/c$ 
as shown by \eref{eOmegak} and \eref{eOmegakinf}. For a finite $D$, the frequencies 
$\omega_{qq'}=\epsilon_q-\epsilon_{q'}=c(q-q')$ are not exactly equal to multiples of 
$\pi c/L$ but most of them are close to these values, see Fig.~\ref{fig:spectre}. As it is 
clear from equation \eref{eqq}, the spacing between consecutive frequencies 
$\omega_{qq'}$ varies and goes to zero for $\omega_{qq'} \rightarrow p\pi c/L$ where 
$p \in \Z$. We see on Fig.~\ref{fig:spectre} that $O_{qq'} \rightarrow 0$ in this limit. 
Time regimes may then exist in which $O(\omega)$ can be approximated as a sum 
of smooth functions with finite support. However, these functions clearly vanish at 
the right end of their support interval but not at the left one. Consequently, their Fourier 
transforms do not vanish fast enough at long times to lead to a relaxation behavior of 
$\langle \hat O (t) \rangle$. We find similar functions $O$ for the energy 
$\langle H_\mathrm{coupling} (t) \rangle$ with the noticeable difference that 
the coefficients $O_{qq'}$ corresponding to the lowest frequencies 
$\omega_{qq'} \simeq \pm 2c(2/DL)^{1/2}$ are dominant in the large $D$ limit, giving 
rise to the sinusoidal behavior discussed at the end of the previous section.

\section{Conclusion}

In this paper, we have studied how changing the thermal contact between 
two quantum systems whose elementary excitations are noninteracting bosons, 
affects these systems. More precisely, our model consists of two one-dimensional 
cavities of equal length coupled to each other by one of their ends. The coupling 
we have considered is thermodynamically small, as shown by the fact that 
the equilibrium total energy and correlation functions are practically independent 
of its strength, and allows to describe from perfectly coupled to completely 
uncoupled systems. We have seen that a sudden change of this thermal contact 
has a considerable impact on the correlation functions of the two cavities. 
They do not return to equilibrium but essentially oscillate with a 
period equal to twice the time required for a signal to propagate from one end of 
a cavity to the other. Their exact time evolution is mainly controlled by the final 
coupling strength which determines the reflection and transmission coefficients 
at the interface between the two systems. We found a similar behavior for 
the energy content of the cavities. For weak coupling between them, it varies 
sinusoidally with a period set by the coupling strength. 
The absence of relaxation behavior is not simply a consequence of the fact 
that the considered Hamiltonian is quadratic. The Joule expansion of a perfect 
quantum gas has, for example, been obtained in Ref.~\cite{PRL}. As discussed 
in the previous section, this difference in behavior comes from qualitative 
differences in the elementary excitation energy spectrum. Clearly, the simple 
dispersion relation of our one-dimensional continuous field model, plays 
a crucial role in the temporal quasiperiodicity obtained. It would be thus interesting 
to study higher-dimensional and lattice free field systems. Are interactions between 
the bosonic excitations necessary to ensure the stability of the thermal equilibrium ? 
 
\appendix

\section{Coupling energy in limiting cases \label{AA}}

To evaluate the energy \eref{HT} for vanishing and large $D$, we make use of 
\begin{equation}
\sum_p F(p) e^{ipx} = \sum_p \int dq F(q) e^{iq(x-2p\pi)} \label{rel}
\end{equation}
where $p$ runs over all integers and $F$ is any function. 

With this relation, we obtain, for large $D$, 
\begin{equation}
\langle H \rangle_T = \frac{T}{2} + \frac{4}{\pi c} LT^2 G(2T\Lambda/c,0) 
+ \frac{8}{\pi c} LT^2 \sum_{p>0} G(2T\Lambda/c,4pTL/c) \label{HTA}
\end{equation}
where
\begin{equation}
G(\epsilon,x)= \mathrm{Re} \int_0^\infty dk \frac{k}{\tanh k} e^{-k\epsilon + ik x} . 
\label{G}
\end{equation}
Using the expansion $\tanh^{-1} k=1+2\sum_{p>0} \exp(-2kp)$ and \eref{rel}, 
we find, in the limit $\epsilon \rightarrow 0$, 
$G(\epsilon,x) = -\pi^2 [2\sinh(\pi x/2)]^{-2}$ 
for $x \ne 0$, and $G(\epsilon,0)=\epsilon^{-2}+\pi^2/12$. Replacing 
these expressions into \eref{HTA} leads to \eref{HTint} with \eref{Einf}. 

For $D=0$, the wavenumbers of the odd eigenmodes are $q=(p+1/2)\pi/L$ 
whereas $q=p\pi/L$ for the even ones. The contributions of the odd and even 
eigenmodes are then different. We obtain
\begin{equation}
E(0,cT/L) = - \frac{2}{\pi c} LT^2 \sum_{p>0}  u_p 
- \frac{2}{\pi c} LT^2 \sum_{p>0} (-1)^p u_p
\end{equation}
where $u_p=\sinh(2p\pi TL/c)^{-2}$, which gives \eref{E0}. 

\section{Time-dependence of the field operators \label{AB}}

The polynomials $P_m^+$ and $P_m^-$  defined after \eref{Gamma} 
are given by
\begin{eqnarray}
P_m^+ (\tau) &=& P_{m-1}^- (\tau)  = \frac{1}{\pi D} \oint_{\cal C} dz F(z) 
\label{Pm} \\
F(z) &=& \frac{ze^{i(z-2i)\tau}}{Dk-z} \left( \frac{z+2i}{z-2i} \right)^m 
\left[ (z-2i)e^{2izL/D} -z-2i \right]^{-1}. \nonumber
\end{eqnarray} 
The above integral is determined by the residue of the function $F$ 
at $z=2i$. We write
\begin{eqnarray}
F(2i+\epsilon) &=& \sum_{p \ge 0} \frac{(i\tau)^p}{p!} \epsilon^{p-m}
\frac{(2i+\epsilon)  ( 4i+\epsilon )^m }{Dk-2i-\epsilon}
\left[ \epsilon e^{(2i\epsilon-4)L/D} -4i -\epsilon \right]^{-1} \nonumber \\
&=& \sum_{p,r \ge 0} \frac{(i\tau)^p}{p!} A_r \epsilon^{p+r-m} .
\end{eqnarray} 
With these notations, \eref{Pm} becomes
\begin{equation}
P_m^+ (\tau) = -\frac{2i}{D} \sum_{p=0}^{m-1}  \frac{(i\tau)^p}{p!} A_{m-p-1}
\end{equation} 
for $m>1$ ($P_0^+=0$). We find, for example, 
\begin{eqnarray}
P_1^+ (\tau) &=& -\frac{4}{D} \frac{1}{Dk-2i} \\
P_2^+ (\tau) &=& \frac{16}{iD} \frac{1}{Dk-2i} \left[ \frac{1}{Dk-2i} 
-\frac{i}{4} \left( e^{-4L/D} +3 \right) + i \tau \right] .
\end{eqnarray}

\end{document}